\documentclass[aps,prd,twocolumn,amsmath,superscriptaddress,amssymb,showpacs,floatfix,nofootinbib,longbibliography]{revtex4-1}

\usepackage[colorinlistoftodos]{todonotes}

\usepackage{graphicx}
\usepackage{dcolumn}
\usepackage{xcolor}
\usepackage{bm}
\usepackage{natbib}
\usepackage{calligra}
\usepackage[T1]{fontenc}
\usepackage{egothic}
\usepackage[T1]{fontenc}
\newfont{\rsfsten}{rsfs10 scaled 1200}
\newfont{\rsfsseven}{rsfs10 scaled 1200}
\newfont{\rsfsfive}{rsfs10 scaled 1200}
\usepackage{epsfig}
\usepackage{units}
\usepackage[utf8]{inputenc}
\usepackage{float}
\usepackage[section]{placeins}

\newcommand{\mnras}{{Mon.~Not.~R.~Astron.~Soc.}}

\newcommand{\jcap}{{Journal~of~Cosmology~and~Astroparticle~Physics}}

\newcommand{\be}{\begin{equation}}
\newcommand{\ee}{\end{equation}}
\newcommand{\bea}{\begin{eqnarray}}
\newcommand{\eea}{\end{eqnarray}}

\newcommand{\barHet}{{}^{3}\overline{\textrm{He}}}
\newcommand{\barHef}{{}^{4}\overline{\textrm{He}}}

\def\lsim{\mathrel{\raise.3ex\hbox{$<$\kern-.75em\lower1ex\hbox{$\sim$}}}}
\def\gsim{\mathrel{\raise.3ex\hbox{$>$\kern-.75em\lower1ex\hbox{$\sim$}}}}

\begin{document}

\vskip 0.2in
\hspace{13cm}\parbox{5cm}{FERMILAB-PUB-20-021-A}
\hspace{13cm}
\vspace{0.6cm}


\title{Anti-Deuterons and Anti-Helium Nuclei from Annihilating Dark Matter}

\author{Ilias Cholis}
\email{cholis@oakland.edu, ORCID: orcid.org/0000-0002-3805-6478}
\affiliation{Department of Physics, Oakland University, Rochester, Michigan, 48309, USA}
\author{Tim Linden}
\email{linden.70@osu.edu, ORCID: orcid.org/0000-0001-9888-0971}
\affiliation{Stockholm University and the Oskar Klein Centre, Stockholm, Sweden}
\affiliation{Center for Cosmology and AstroParticle Physics (CCAPP) and Department of Physics, The Ohio State University Columbus, Ohio, 43210}
\author{Dan Hooper}
\email{dhooper@fnal.gov,  ORCID: orcid.org/0000-0001-8837-4127}
\affiliation{Theoretical Astrophysics Group, Fermi National Accelerator Laboratory, Batavia, Illinois, 60510, USA}
\affiliation{Department of Astronomy and Astrophysics and the Kavli Institute for Cosmological Physics (KICP), University of Chicago, Chicago, Illinois, 60637, USA}

\date{\today}

\begin{abstract}
Recent studies of the cosmic-ray antiproton-to-proton ratio have identified an excess of $\sim$\mbox{10--20~GeV} antiprotons relative to the predictions of standard astrophysical models. Intriguingly, the properties of this excess are consistent with the same range of dark matter models that can account for the long-standing excess of $\gamma$-rays observed from the Galactic Center. Such dark matter candidates can also produce significant fluxes of anti-deuterium and anti-helium nuclei. Here we study the production and transport of such particles, both from astrophysical processes as well as from dark matter annihilation. Importantly, in the case of \textit{AMS-02}, we find that Alfv{\'e}nic reacceleration (i.e., diffusion in momentum space) can boost the expected number of $\bar{\rm d}$ and ${}^{3}\overline{\textrm{He}}$ events from annihilating dark matter by an order of magnitude or more. For relatively large values of the Alfv{\'e}n speed, and for dark matter candidates that are capable of producing the antiproton and $\gamma$-ray excesses, we expect annihilations to produce a few anti-deuteron events and about one anti-helium event in six years of \textit{AMS-02} data. This is particularly interesting in light of recent reports from the \textit{AMS-02} Collaboration describing the detection of a number of anti-helium candidate events.

\end{abstract}

\maketitle

Measurements of high-energy antimatter in the cosmic-ray spectrum provide a powerful probe of new physics, including the annihilation or decay of dark matter particles in the halo of the Milky Way~\cite{Bergstrom:1999jc, Hooper:2003ad, Profumo:2004ty, Bringmann:2006im, Pato:2010ih}. An excess of $\sim$\mbox{10--20} GeV cosmic-ray antiprotons~\cite{Hooper:2014ysa, Cirelli:2014lwa, Bringmann:2014lpa, Cuoco:2016eej, Cui:2016ppb} has been identified in data from \textit{AMS-02} \cite{Aguilar:2016kjl} (and \textit{PAMELA}~\cite{Adriani:2010rc}), with characteristics that are consistent with the annihilation of $\sim$\mbox{50--90} GeV dark matter particles with a cross section of $\langle \sigma v\rangle \simeq (1-9) \times 10^{-26}$ cm$^{3}$/s~\cite{Cholis:2019ejx} (for the case of annihilations to $b\bar{b}$; for other dark matter scenarios consistent with this signal, see Refs.~\cite{Cholis:2019ejx,Hooper:2019xss}). This excess appears to be statistically significant ($>$3.3$\sigma$), and robust to systematic uncertainties associated with the antiproton production cross section, the propagation of cosmic rays through the interstellar medium (ISM), and the time-, charge- and energy-dependent effects of solar modulation~\cite{Cuoco:2019kuu, Cholis:2019ejx} (see, however~\cite{Reinert:2017aga, Boudaud:2019efq}). Intriguingly, the range of dark matter models favored by the antiproton data is also consistent with that required to explain the $\gamma$-ray excess that has been observed from the direction of the Galactic Center~\cite{Hooper:2010mq, Hooper:2011ti,  Abazajian:2012pn, Gordon:2013vta, Daylan:2014rsa,Calore:2014xka,TheFermi-LAT:2015kwa}.

In 2015, two groups analyzed small-scale power in the $\gamma$-ray data and argued that the $\gamma$-ray excess is likely generated by a large population of point sources (such as millisecond pulsars)~\cite{Bartels:2015aea, Lee:2015fea}. More recent work, however, has shown the interpretation of these results to be problematic~\cite{Leane:2019uhc,Zhong:2019ycb}. In particular, Ref.~\cite{Leane:2019uhc} demonstrated that one class of algorithms is systematically biased towards pulsar models and is unable to recover true dark matter signals that are injected into the data (in stark contrast with the results of a recent study limited to the case of mock data~\cite{Chang:2019ars}). Recent work by Ref.~\cite{Zhong:2019ycb} has gone farther, placing constraints on the luminosity function of any point source population that are in strong tension with millisecond pulsar models~\cite{Cholis:2014lta,Hooper:2016rap,Hooper:2015jlu,Ploeg:2017vai,Bartels:2018xom}.

In addition to $\gamma$-rays and antiprotons, dark matter annihilations can produce potentially detectable fluxes of heavier anti-nuclei, including anti-deuterons and anti-helium~\cite{Donato:1999gy}. As kinematic considerations strongly suppress the production of heavy anti-nuclei in astrophysical processes, the detection of such particles 
could constitute a smoking-gun for dark matter annihilation. Intriguingly, the \textit{AMS-02} Collaboration has reported preliminary evidence of $\mathcal{O}(10)$ candidate anti-helium events~\cite{AMSLaPalma}. Such a rate would be very surprising, as it would significantly exceed that predicted from standard astrophysical processes or from annihilating dark matter~\cite{Carlson:2014ssa, Cirelli:2014qia}, and no other plausible means of producing so much high-energy anti-helium has been identified~\cite{Poulin:2018wzu} (see also Refs.~\cite{Aramaki:2015laa, Blum:2017qnn, Korsmeier:2017xzj, Li:2018dxj}). For example, while several groups have found that the uncertainties associated with the anti-nuclei production cross sections could substantially increase the number of anti-helium events from dark matter annihilations~\cite{Carlson:2014ssa, Cirelli:2014qia}, these rates still lie well below those required to explain the preliminary results from \textit{AMS-02}.

In this \emph{letter}, we investigate how variations in the cosmic-ray transport model could impact the local spectrum of anti-deuterium and anti-helium. Most significantly, we find that diffusive reacceleration (also known as Alfv{\'e}nic reacceleration or diffusion in momentum space) could dramatically increase the number of anti-deuterium and anti-helium events predicted to be observed by \textit{AMS-02} in annihilating dark matter scenarios. Furthermore, this effect is more pronounced for anti-helium than for anti-deuterium, potentially helping to explain the unexpectedly large number of anti-helium candidate events. In models where the dark matter's mass, annihilation cross section and final state are chosen to fit the antiproton and $\gamma$-ray excesses, and for a relatively large Alfv{\'e}n speed of $v_A \sim 60$ km/s, we expect \textit{AMS-02} to detect roughly one $\barHet$ event and a few $\bar{\rm d}$ events (in 6 years of data).



Throughout this study, we will consider two mechanisms for the production of cosmic-ray antiprotons, anti-deuterons and anti-helium nuclei. First, antimatter can be produced through the collisions of primary cosmic rays with interstellar gas. The flux and spectrum of this contribution depend on the primary cosmic-ray spectrum and on the average quantity of gas that the cosmic rays encounter before escaping the Milky Way.\footnote{We adopt a spectrum and spatial distribution of injected primary cosmic rays that provides a good fit to the measured primary-to-secondary ratios. More specifically, the injected spectra are described by a broken power-law in rigidity with an index of 1.9 (\mbox{2.38--2.45}), below (above) a break of 11.7~GV for all cosmic-ray species~\cite{Cholis:2015gna, Cholis:2017qlb, Cholis:2019ejx}.} Second, antimatter can be produced through the annihilation of dark matter particles, with a spectrum that depends on the distribution of dark matter, as well as on the characteristics of the dark matter candidate itself. For this case, we adopt an Navarro-Frenk-White (NFW) profile~\cite{Navarro:1995iw} with a local density of 0.4~GeV/cm$^3$~\cite{Catena:2009mf, Salucci:2010qr} and a scale radius of 20~kpc. In both cases, we calculate the spectrum of anti-nuclei that is injected into the ISM using a ``coalescence'' model~\cite{Chardonnet:1997dv}, in which two anti-nucleons are predicted to fuse into a common nucleus if the difference between their relative momenta is smaller than the coalescence momentum, $p_0$ (for details, see the Supplementary Material provided in the Appendix).


To model the transport of cosmic rays through the ISM, we use the publicly available code \texttt{Galprop} v56~\cite{GALPROPSite, galprop}, which accounts for the effects of cosmic-ray diffusion, convection, diffusive reacceleration and fragmentation, as well as energy losses from ionization and Coulomb interactions, synchrotron, bremsstrahlung and inverse Compton scattering emission. 
%
%
This transport model assumes diffusion to be isotropic and homogeneous within a cylindrical zone centered at the Galactic Center.  We adopt a diffusion coefficient of the form $D_{xx}(R) = \beta D_{0} (R/4 GV)^{\delta}$, where $\beta \equiv v/c$ and $\delta \sim \mbox{0.3--0.5}$~\cite{Trotta:2010mx} is the diffusion index associated with magnetohydrodynamic turbulence in the ISM. We additionally allow particles to be propelled out of the plane by convective winds, with a speed that is zero at the plane and that increases at larger heights as:
\begin{equation}
v_{c} = \frac{dv_{c}}{d|z|} |z|.
\label{eq:Conv}
\end{equation}

The spatial diffusion of cosmic rays is the result of scattering on magnetic turbulence. In addition, cosmic rays experience diffusive reacceleration (or Alfv{\'e}nic reacceleration) due to the resonant interaction of charged particles of a given gyroradius with the corresponding Alf{\'e}n modes of the turbulent medium. This is manifest as diffusion in momentum space with the following coefficient~\cite{1994ApJ...431..705S}:
\begin{equation}
D_{pp} \propto \frac{R^{2}v_{A}^{2}}{D_{xx}(R)},
\label{eq:DiffReAcc}
\end{equation}
where the Alfv$\acute{\textrm{e}}$n speed, $v_{A}$, is the speed that hydromagnetic waves propagate through the ISM plasma.

We begin our analysis using ISM Models I,II, and III from Ref.~\cite{Cholis:2019ejx}, and then consider alterations to these reference scenarios. While diffusion, convection, and diffusive reacceleration each play significant roles in determining the spectra of cosmic-ray anti-nuclei, we find that the very strong constraints on the diffusion coefficient (from measurements of the B/C ratio and Voyager measurements of the low-energy cosmic-ray flux), strongly constrain the combined values of $D_0$ and $\delta$. On the other hand, the current data allow for much larger variations in in the parameters that describe the effects of convection and diffusive reacceleration, making them the dominant sources of uncertainty in our prediction for the resulting spectra of cosmic-ray anti-nuclei. To account for the uncertainties associated with the effects of solar modulation, we use the model described in Ref.~\cite{Cholis:2015gna}. Our approach is the same as that adopted in Refs.~\cite{Cholis:2015gna, Cholis:2017qlb, Cholis:2019ejx}, accounting for measurements of the magnitude of the heliospheric magnetic field at Earth from \textit{ACE}~\cite{ACESite} and of the morphology of this field from the Wilcox Solar Observatory~\cite{WSOSite}.

\begin{figure}
\hspace{-0.0in}
\includegraphics[width=3.45in,angle=0]{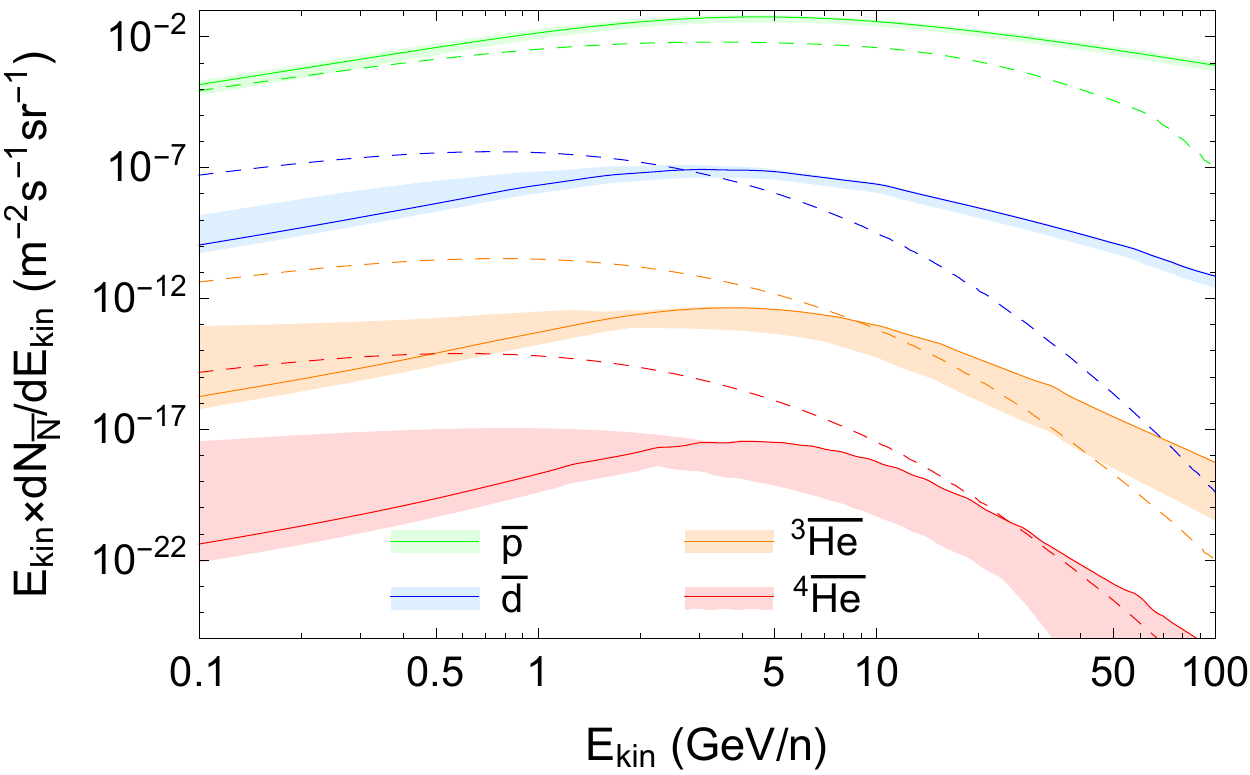}
\vskip -0.1in
\caption{The spectrum of cosmic-ray $\bar{\rm p}$ (green), $\bar{\rm d}$ (blue), $\barHet$ (orange) and 
$\barHef$ (red) predicted from standard astrophysical production (solid curves), along with the uncertainty associated with this prediction (bands), for the case of ISM Model I. The dashed curves are the central prediction for the anti-nuclei spectra from an annihilating dark matter model that is capable of producing the antiproton and $\gamma$-ray excesses ($m_{\chi}=67$ GeV, $\sigma v =2\times 10^{-26}$ cm$^3/s$, $\chi \chi \rightarrow b\bar{b}$). Note that diffusive reacceleration can lead to non-zero anti-nuclei fluxes from annihilating dark matter well above the maximum injected energy of such particles.}
\label{fig:AntiNucleiFluxes}
\end{figure}

 \begin{figure*}[!]
\includegraphics[width=3.41in,angle=0]{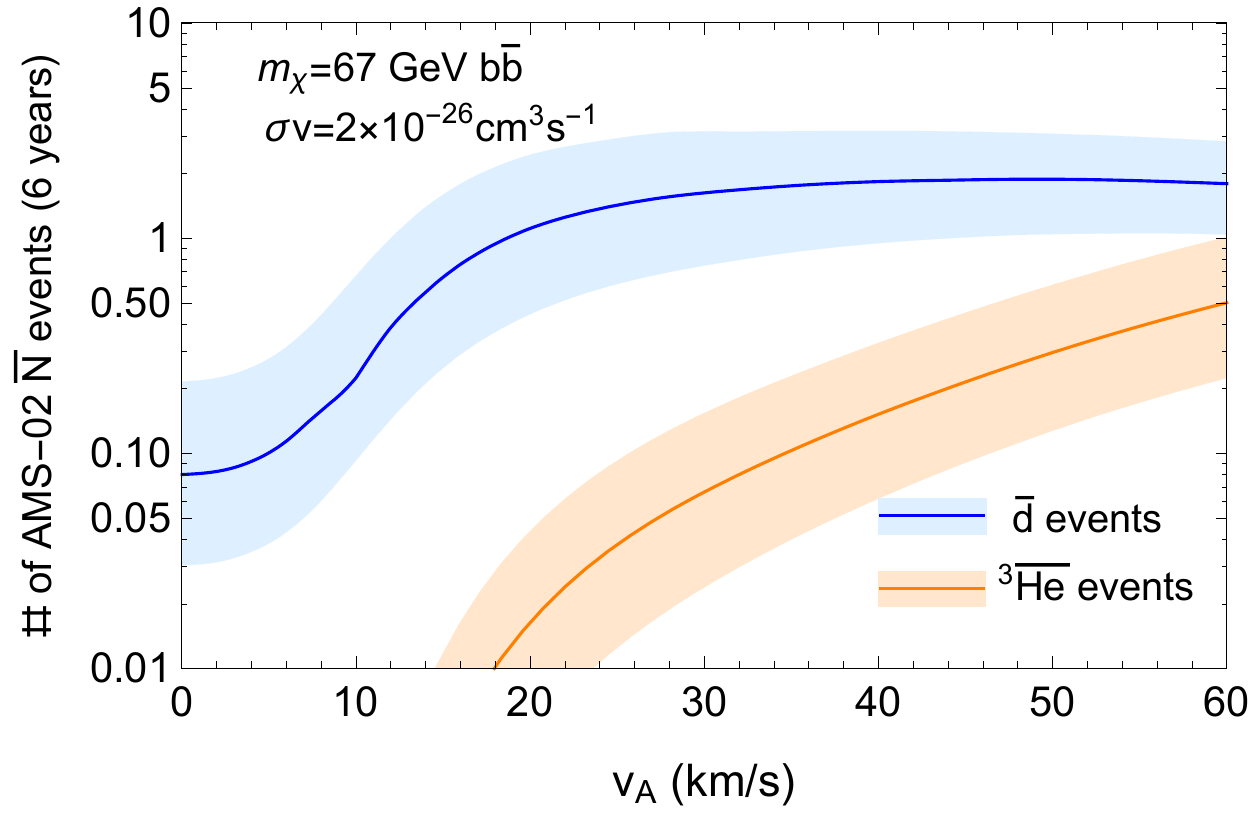}
\includegraphics[width=3.41in,angle=0]{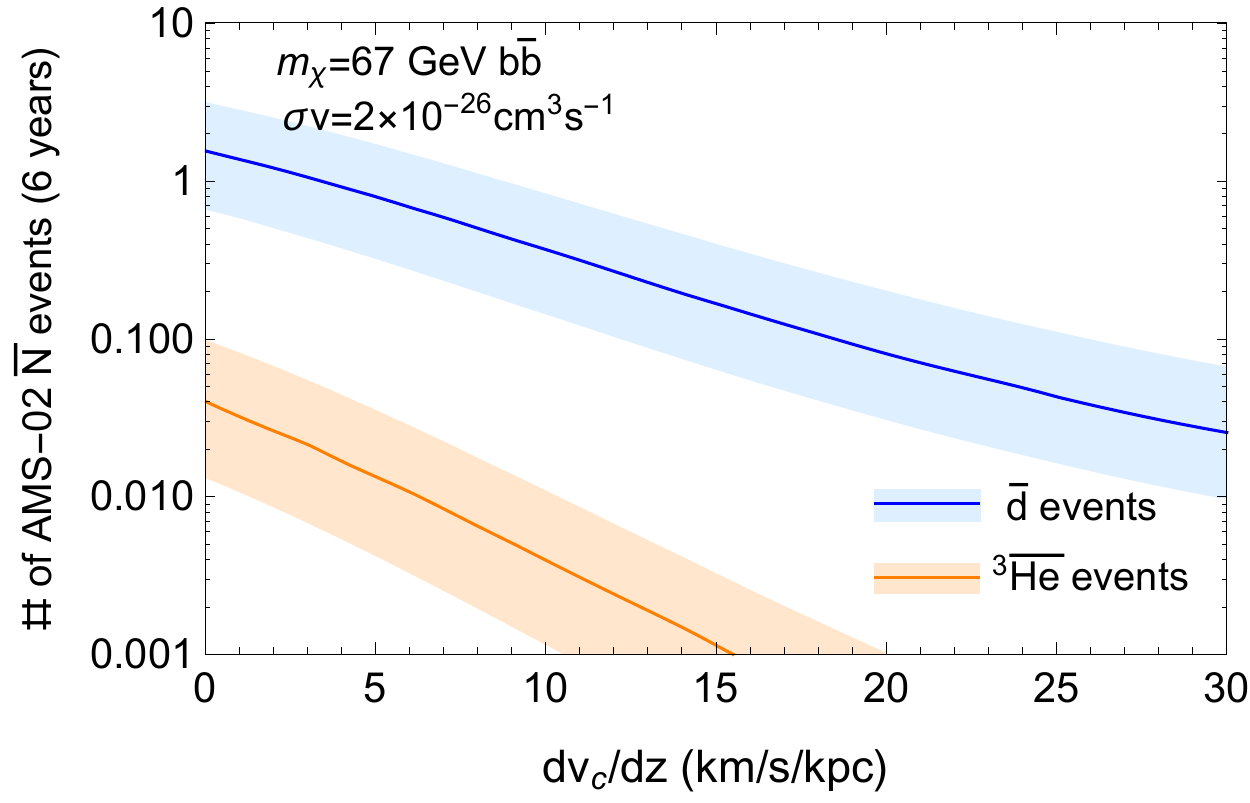}
\vskip -0.1in
\caption{The number of $\bar{\rm d}$ and $\barHet$ events predicted to be observed by \textit{AMS-02} in six years of data from annihilating dark matter ($m_{\chi}=67$ GeV, $\sigma v =2\times 10^{-26}$ cm$^3/s$, $\chi \chi \rightarrow b\bar{b}$). In the left (right) frame, we vary the Alfv{\'e}n speed (convection velocity gradient) while keeping all of the other propagation parameters set to those of ISM Model I~\cite{Cholis:2019ejx}. The bands represent the uncertainties associated with the coalescence momentum and solar modulation.}
\label{fig:Convection_and_Reacceleration}
\end{figure*} 

In Fig.~\ref{fig:AntiNucleiFluxes}, we plot the spectrum of $\bar{\rm p}$, $\bar{\rm d}$, $\barHet$ and $\barHef$ from standard astrophysical production, and compare this to the contributions predicted from annihilating dark matter. In particular, we consider a dark matter model that is capable of producing the observed features of the antiproton and $\gamma$-ray excesses ($m_{\chi}=67$ GeV, $\sigma v=2\times 10^{-26}$ cm$^3/s$ and $\chi\chi \rightarrow b\bar{b}$)~\cite{Calore:2014xka,Cuoco:2016eej, Cui:2016ppb, Cuoco:2017rxb, Cholis:2019ejx}. The solid curves represent the central value of the astrophysical predictions, while the surrounding bands reflect the total uncertainty, adopting the assumptions described in the Supplementary Material
(and adopting $p_{0} = 261$ MeV for $\barHet$ and $\barHef$). The dashed lines depict our central prediction from the selected annihilating dark matter model.

We note that the contribution from dark matter dominates the spectrum of anti-nuclei at low-energies. This is due to the fact that dark matter annihilation occurs in the laboratory frame, while astrophysical secondary production necessarily occurs in a boosted frame. Furthermore, the dark matter contribution becomes increasingly dominant for more massive anti-nuclei, due to the increasing kinematic suppression of secondary production mechanisms. In Table~\ref{tab:fitEvents}, we show the number of $\bar{\rm d}$, $\barHet$ and $\barHef$ events that we predict \textit{AMS-02} will observe with six years of data, from astrophysical secondary production and from dark matter annihilation. To calculate these rates, we combine the spectra shown in Fig.~\ref{fig:AntiNucleiFluxes} with the reported sensitivity of \textit{AMS-02}~\cite{Aramaki:2015pii}.

\begin{table}[!]
    \begin{tabular}{ccccc}
         \hline
                & Astro  & DM (I) &  DM (II) &  DM (III) \\
            \hline \hline
             $\bar{\rm d}$&\footnotesize{0.02-0.1}&  \footnotesize{0.6-3.0}& \footnotesize{0.4-2} & \footnotesize{0.3-1.5} \\
             $\barHet$&\footnotesize{(0.3-3)$\times 10^{-3}$}&  \footnotesize{0.01-0.1} & \footnotesize{(0.6-6)$\times 10^{-2}$} & \footnotesize{(0.5-5)$\times 10^{-2}$} \\
             $\barHef$&\footnotesize{(0.06-6)$\times 10^{-9}$}& \footnotesize{(0.2-5)$\times10^{-4}$} & \footnotesize{(0.8-15)$\times 10^{-6}$} & \footnotesize{(0.6-12)$\times 10^{-6}$} \\
             \hline \hline 
        \end{tabular}
       \caption{The number of $\bar{\rm d}$, $\barHet$ and $\barHef$ events that \textit{AMS-02} is expected to observe with six years of data from astrophysical secondary production (``Astro'') and from dark matter annihilation. For the case of dark matter, we adopt a model that is capable of producing the antiproton and $\gamma$-ray excesses ($m_{\chi}=67$ GeV, $\sigma v =2\times 10^{-26}$ cm$^3/s$, $\chi \chi \rightarrow b\bar{b}$), and show rates using three ISM transport models (Models I, II and III of Ref.~\cite{Cholis:2019ejx}). For the case of astrophysical secondary production we have marginalized over these three models. The ranges shown include the uncertainties associated with in the coalescence momenta, the proton-proton cross section, and the effects of solar modulation.}
    \label{tab:fitEvents}
\end{table}

At this point, we would like to emphasize two aspects of our results. First, the rate of anti-nuclei events from dark matter consistently exceeds that predicted from astrophysical production. 
And second, the number of events from secondary production is not significantly affected by the choice of ISM Model, while the dark matter flux can change by several orders of magnitude in different ISM Models. This is due to the significant effect of cosmic-ray propagation on the spectrum of anti-nuclei from dark matter annihilation. As we will show, this is the key result of this letter.

In Fig.~\ref{fig:Convection_and_Reacceleration}, we illustrate the impact of diffusive reacceleration and convection on the number of anti-nuclei events from dark matter predicted to be observed by \textit{AMS-02}. Here, we have fixed all of the propagation parameters to those of ISM Model I, with the exception of the Alfv{\'e}n speed and convection velocity, which are varied in the left and right frames, respectively (the default values in ISM Model I are $v_A$~=~24~km/s and $dv_{c}/d|z| = 1$ km/s/kpc). The shaded bands in this figure represent the uncertainties associated with coalescence and solar modulation. 

While the predicted number of anti-deuteron events is roughly flat for values of $v_A$ above $\sim$20~km/s, the number of anti-helium events is highly sensitive to this quantity. For large values of $v_A$, it is entirely plausible that \textit{AMS-02} could detect on the order of one $^3\overline{\textrm{He}}$ event over six years of operation. In contrast, increasing the convection velocity has the effect of suppressing the number of anti-deuteron and anti-helium events observed by \textit{AMS-02}.

\begin{figure*}
\hspace{-0.0in}
\includegraphics[width=3.45in,angle=0]{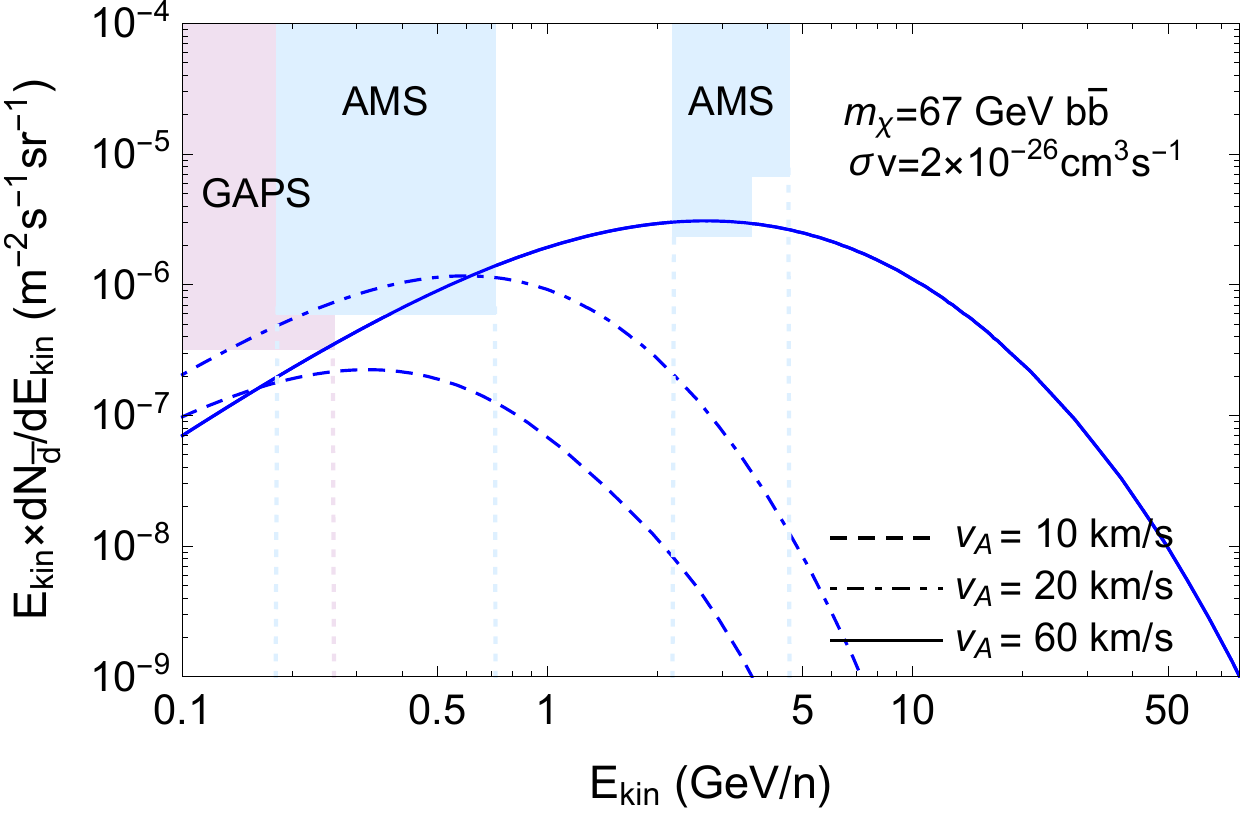}
\includegraphics[width=3.45in,angle=0]{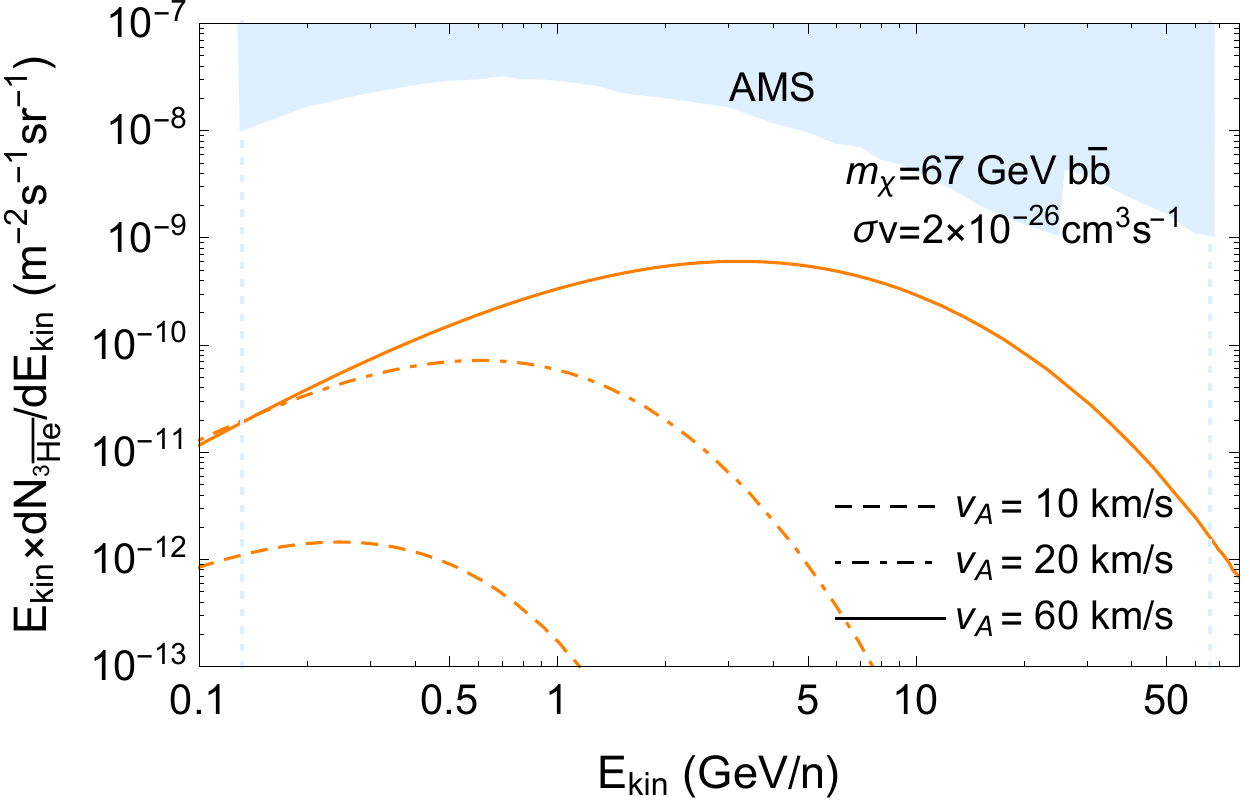}
\vskip -0.1in
\caption{The spectrum of cosmic-ray anti-deuterons (left) and anti-helium nuclei (right) from annihilating dark matter ($m_{\chi}=67$ GeV, $\sigma v =2\times 10^{-26}$ cm$^3/s$, $\chi \chi \rightarrow b\bar{b}$), for three values of the Alfv{\'e}n speed. This is compared to the sensitivities of the \textit{AMS-02}~\cite{Aramaki:2015pii} (blue) and GAPS~\cite{Aramaki:2015laa} (purple) experiments.}
\label{fig:change}
\end{figure*}

To understand the dependence of these event rates on the Alfv{\'e}n speed, it is important to appreciate that \textit{AMS-02} is sensitive to anti-nuclei only across a limited range of energies. As illustrated in Fig.~\ref{fig:change}, \textit{AMS-02} reports sensitive to anti-deuterons only in the ranges of \mbox{0.18--0.72} and \mbox{2.2--4.6} GeV/n. As dark matter annihilations are predicted to produces a large flux anti-deuterons below this range of energies, even a modest amount of diffusive reacceleration can quite dramatically increase the rate at which such particles are ultimately detected by \textit{AMS-02}. The rate of $\barHet$ events is even more sensitive to the Alfv{\'e}n speed due to the higher energy range across which \textit{AMS-02} can detect and identify such particles~\cite{Kounine}. Convective winds, on the other hand, have the effect of reducing the local flux of anti-nuclei. This is particularly important at low energies, where diffusion is less efficient.

Relative to that from annihilating dark matter, the rate of anti-nuclei events from astrophysical secondary production is far less sensitive to the values of the Alfv{\'e}n speed or convection velocity. This is due to the kinematics associated with the two processes. In particular, the secondary production of $\bar{\rm d}$ ($\barHet$) requires the primary cosmic ray to have a kinetic energy in excess of 17~$m_p$ (31~$m_p$), and thus such particles are invariably highly boosted~\cite{Donato:1999gy}. In contrast, dark matter annihilations occur in the center-of-mass frame.

\begin{figure}[ht!]
\includegraphics[width=3.41in,angle=0]{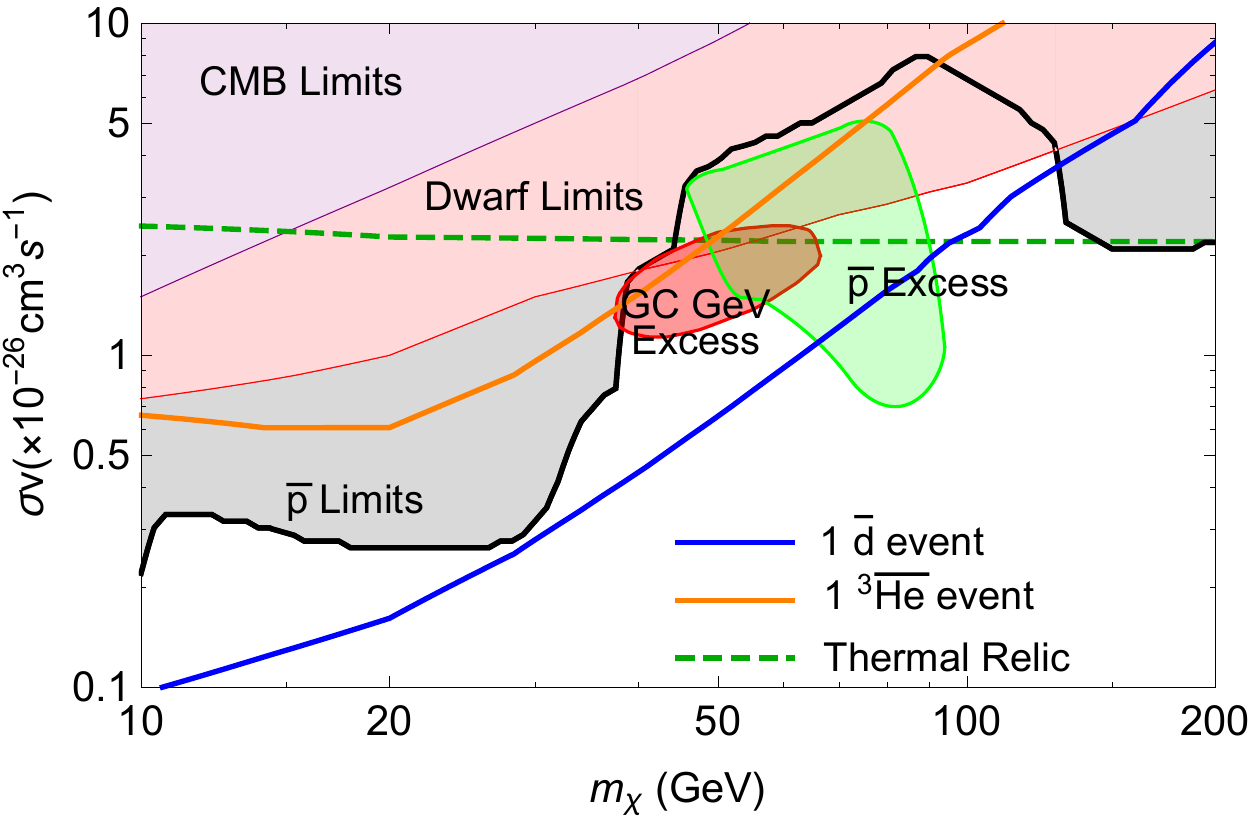}
\vskip -0.1in
\caption{The regions of the dark matter parameter space (for the case of $\chi \chi \rightarrow b \bar{b}$) in which one would expect \textit{AMS-02} to observe one $\bar{\rm d}$ or one $\barHet$ event in six years of data. Also shown are the regions that are consistent with the observed characteristics of the antiproton~\cite{Cholis:2019ejx} and $\gamma$-ray~\cite{Calore:2014xka} excesses, as well as the constraints derived from gamma-ray observations of dwarf galaxies~\cite{Fermi-LAT:2016uux}, the cosmic microwave background (CMB)~\cite{Aghanim:2018eyx}, and the cosmic-ray antiproton-to-proton ratio~\cite{Cholis:2019ejx}. Here we have adopted a coalescence momentum of $p_{0} = 160$ MeV, an Alfv{\'e}n speed of $v_A=60$~km/s, and ISM Model I.}
\label{fig:DM_limits_andExcess}
\end{figure}

Finally, we show in Fig.~\ref{fig:DM_limits_andExcess} the regions of the dark matter parameter space in which one would expect \textit{AMS-02} to observe one $\bar{\rm d}$ or one $\barHet$ event, and compare this to the regions that are consistent with the observed characteristics of the antiproton and $\gamma$-ray excesses, as well as the constraints derived from gamma-ray observations of dwarf galaxies~\cite{Fermi-LAT:2016uux} and the cosmic microwave background (CMB)~\cite{Aghanim:2018eyx}. For the parameter values considered (ISM Model I, $v_A=60$~km/s, $p_0=160$~MeV), the regions favored by the observed excesses are predicted to result in roughly $\sim$~1 $\barHet$ event and a few $\bar{\rm d}$ events (in 6 years of \textit{AMS-02} data).

In summary, even after accounting for the uncertainties associated with the proton-proton cross section, the coalescence model, the injection into and transport through the ISM, and solar modulation, we find that astrophysical secondary production cannot produce a flux of anti-deuterons or anti-helium nuclei that would be detectable by \textit{AMS-02}. On the other hand, dark matter that annihilates to hadronic final states (or to particles that decay hadronically~\cite{Hooper:2019xss}) could potentially produce a detectable flux of such particles. We emphasize that the rate of anti-nuclei events from dark matter annihilations depends very sensitively on the impact of diffusive reacceleration. In particular, we have demonstrated that by increasing the Alfv{\'e}n speed from 10 km/s to 60 km/s, for example, one could increase the predicted rate of anti-deuteron (anti-helium) events from dark matter by more than an order of magnitude (two orders of magnitude); see Fig.~\ref{fig:Convection_and_Reacceleration}. To our knowledge, this fact has not been previously discussed in the literature.

The \textit{AMS-02} Collaboration has recently reported the tentative observation of $\mathcal{O}(10)$ candidate $\barHet$ events. Needless to say, this would be an incredibly exciting result if confirmed. While our model does not naively predict such a large number of $\barHet$ events, the results presented here provide an important path forward toward understanding the uncertainties that must be at play in order to account for such a large signal. More specifically, our results indicate that if dark matter is responsible for producing  more than a few $\barHet$ events at \textit{AMS-02}, the coalescence momentum for anti-helium must significantly exceed the constraints presented in Ref.~\cite{Schael:2006fd}, and the average Alfv{\'e}n speed must significantly exceed the best-fit values of standard {\tt Galprop} models ($\sim$\mbox{20~km/s}). Notably, both of these quantities can be independently probed with new data (the degeneracy between the Alfv{\'e}n speed and other propagation parameters will be the subject of future work). 
We note that this antinuclei 
flux is produced at regions of high dark matter density as the inner two kiloparsecs of the Milky-Way or close-by subshalos. Thus, the values of high Alfv{\'e}n speed do not need to represent the entirety of the ISM just the regions with high dark matter density \footnote{Ref. \cite{Drury:2016ubm}, placed constraints on the maximum averaged Alfv{\'e}n speed (as low as 30 km/s) in the Milky Way's ISM based on cosmic-ray power requirements. However, those are sensitive to the exact ISM diffusion assumptions. Also Refs. \cite{Orlando:2013ysa, 2013JCAP...03..036D, Orlando:2017mvd} have placed constraints from the associated synchrotron emission of cosmic-ray electrons the strength of which depend on the B-field morphology and magnitude.}.

Lastly, the results presented here are consistent with the possibility that both the $\gamma$-ray excess from the Galactic Center~\cite{Hooper:2010mq,Hooper:2011ti,Abazajian:2012pn, Gordon:2013vta,Daylan:2014rsa,Calore:2014xka,TheFermi-LAT:2015kwa} and the cosmic-ray antiproton-excess  \cite{Cuoco:2016eej, Cuoco:2019kuu, Cholis:2019ejx} could arise from a 50-80 GeV dark matter particle annihilating with a cross section near that predicted for a thermal relic, $\langle \sigma v\rangle \sim 2\times 10^{-26}$ cm$^3/$s. Furthermore, in such a scenario, we expect \textit{AMS-02} to observe up to roughly one anti-helium event and a few anti-deuterons, depending on the values of the Alfv{\'e}n speed and convection velocity. Moreover, such a model remains consistent with current constraints from dwarf spheroidal galaxies and the cosmic microwave background.

\begin{acknowledgments}
                  
We would like to thank Simeon Bird and Marc Kamionkowski for valuable discussions, Peter Reimitz for assistance with the HERWIG code and Michael Korsmeier and Martin Winkler for valuable comments on the manuscript. 
IC is partially supported by the Oakland University Research Committee Faculty Fellowship Award. 
TL is partially supported by the Swedish Research Council under contract 2019-05135, the Swedish National Space Agency under contract 117/19 and the European Research Council under grant 742104.
DH is supported by the US Department of Energy under contract DE-FG02-13ER41958. Fermilab is operated by Fermi Research Alliance, LLC, under Contract No. DE- AC02-07CH11359 with the US Department of Energy. 
\end{acknowledgments}

\bibliography{DM_AMS02_antinuclei}
 
\begin{appendix}

\section{The Injected Spectrum of Cosmic-Ray Anti-Nuclei}

\begin{figure}[t]
\hspace{-0.0in}
\includegraphics[width=3.45in,angle=0]{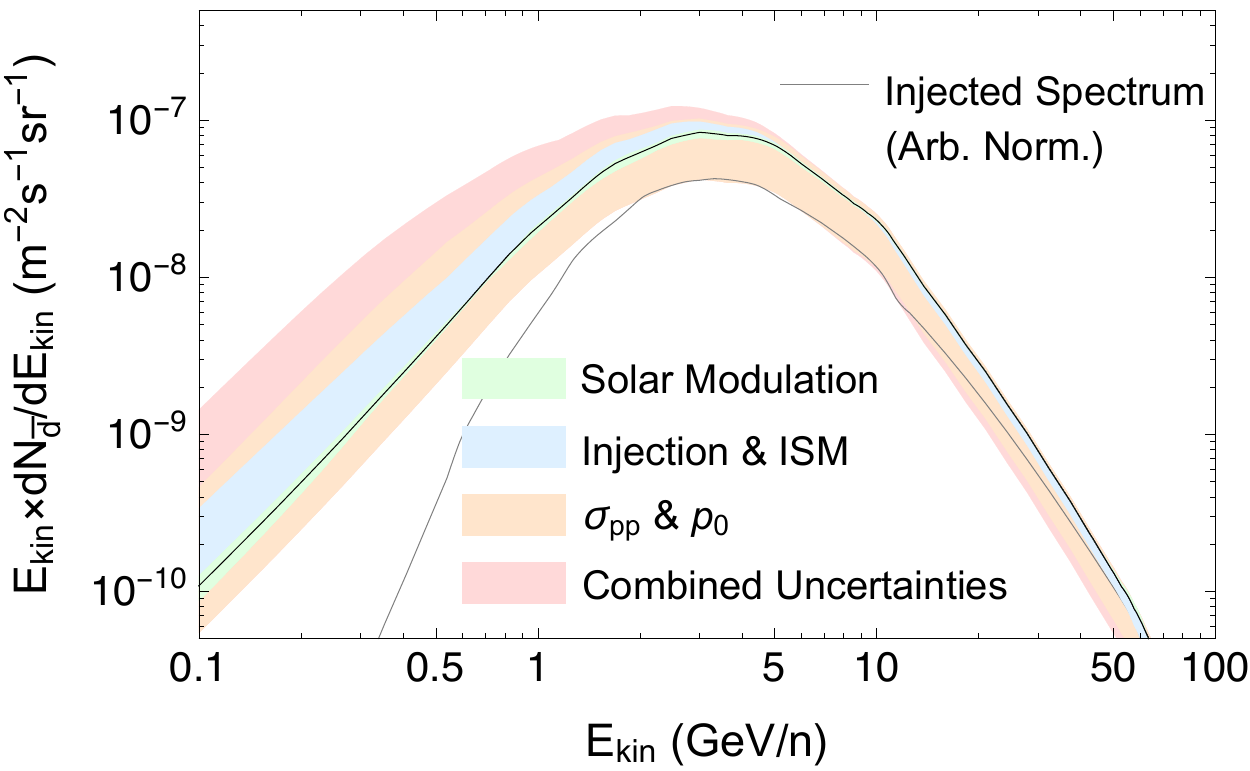}
\vskip -0.1in
\caption{The spectrum of cosmic-ray anti-deuterons predicted from standard astrophysical production, along with the various uncertainties associated with this prediction. The black curve is our central prediction for the case of ISM Model I and the colored bands represent the uncertainties associated with solar modulation (green), the injection and ISM transport model (blue) and the antiproton production cross section and coalesence momentum (orange). The red band depicts the total uncertainty associated with the combination of these factors.}
\label{fig:AntiDeuteronsFluxes}
\end{figure}  

To calculate the spectrum of anti-nuclei produced through dark matter annihilation, we first determine the spectrum of antinucleons using \texttt{PPPC4DMID}~\cite{Cirelli:2010xx}.\footnote{In comparing the output of \texttt{PPPC4DMID} to the that obtained using the Monte Carlo event generators \texttt{PYTHIA}~\cite{Sjostrand:2007gs} and \texttt{HERWIG}~\cite{Corcella:2000bw}, we find that an order one differences can in some cases arise in the $\gamma$-ray and antiproton spectra, resulting from variations in the underlying hadronization and fragmentation algorithms~\cite{Meade:2009rb}. More specifically, we find that \texttt{HERWIG} typically predicts lower fluxes of antiprotons, antideuterons, and anti-helium nuclei than either \texttt{PYTHIA} or \texttt{PPPC4DMID}.} We then model the relevant nuclear physics involved, employing the so-called  ``coalescence'' model~\cite{Chardonnet:1997dv}, in which two antiprotons or antineutrons combine to form a common nucleus if the difference between their relative momenta is smaller than the coalescence momentum, $p_0$. We further simplify this calculation by assuming that the production of a second anti-nucleon is independent of the production probability of the first (that is, there is no correlation between the flux or momentum of the particles in individual collisions)~\cite{Korsmeier:2017xzj}. Simulations of correlated antiparticle production in Monte Carlo event generators have found that this assumption is adequate in the $m_{\chi}\sim$~\mbox{50--100~GeV} range considered here~\cite{Kadastik:2009ts}. Under this assumption, the differential cross sections for $\bar{\rm d}$, $\barHet$  and $\barHef$ production can be written as:
\begin{eqnarray}
E_{A,Z} \frac{d^{3}\sigma_{A,Z}}{dp^{3}_{A,Z}} &=& \frac{m_{A,Z}}
{m_{ p}^{| Z |}m_{ n}^{ A-| Z |}}  
\left ( \frac{1}{\sigma_{pp}} \frac{4 \pi}{3} \frac{p_{0}^{3}}{8} \right )^{A-1} \\ \nonumber
&\times& \left ( E_{\bar{ p}}\frac{d^{3}\sigma_{\bar{ p}}}{dp^{3}_{\bar{ p}}}\right )^{| Z|}
 \left ( E_{\bar{ n}}\frac{d^{3}\sigma_{\bar{ n}}}{dp^{3}_{\bar{ n}}}\right )^{ A-| Z|}, \,\,\,
\label{eq:CSantinuclei}
\end{eqnarray}
where $\sigma_{pp}$ is the total proton-proton cross section and $A$ and $Z$ are the mass and atomic number of the species, respectively. For the case of $dN_{\bar{p}}/dE_{\bar{p}} = dN_{\bar{n}}/dE_{\bar{n}}$, this leads to the following differential spectra of anti-nuclei:
\begin{eqnarray}
\label{eq:CSantinucleiD}
\frac{dN_{\bar{d}}}{dE_{\bar{d}}} &=& \frac{m_{\bar{d}}}{m_{p} m_{n}} 
\frac{4}{3} \frac{p_{0}^{3}} {8 p_{\bar{d}}} \frac{dN_{\bar{p}}}{dE_{\bar{p}}}  
\frac{dN_{\bar{n}}}{dE_{\bar{n}}}, \\
\frac{dN_{\, \barHet}}{dE_{\, \barHet}} &=& \frac{m_{\,  \barHet}}{m_{p}^{2} m_{n}} 
3  \left (\frac{p_{0}^{3}}{8 p_{\, \barHet}} \right)^{2}  \left (\frac{dN_{\bar{p}}}{dE_{\bar{p}}} \right)^{2} 
\frac{dN_{\bar{n}}}{dE_{\bar{n}}}, \nonumber \\
\frac{dN_{\,  \barHef}}{dE_{\,  \barHef}} &=&  \frac{m_{\,  \barHef}}{m_{p}^{2} m_{n}^{2}} 
\frac{4^{4}}{3^{3}} \left (\frac{p_{0}^{3}}{8 p_{\,  \barHef}} \right)^{3} 
\left (\frac{dN_{\bar{p}}}{dE_{\bar{p}}} \right)^{2} 
\left (\frac{dN_{\bar{n}}}{dE_{\bar{n}}} \right)^{2}. \nonumber
\end{eqnarray}

We evaluate the fluxes of secondary $\bar{\rm d}$, $\barHet$ and $\barHef$ for primary cosmic ray species as heavy as silicon, and include interactions with both hydrogen and helium gas. We note that $\barHet$ can be formed both directly, or through the decay of anti-tritium at an equivalent rate. In all cases, we vary the normalization of the total Milky Way gas density from default \texttt{Galprop} values by up to $\pm$10\%.

\begin{figure*}[!]
\vspace{0.1in} 
\includegraphics[width=3.41in,angle=0]{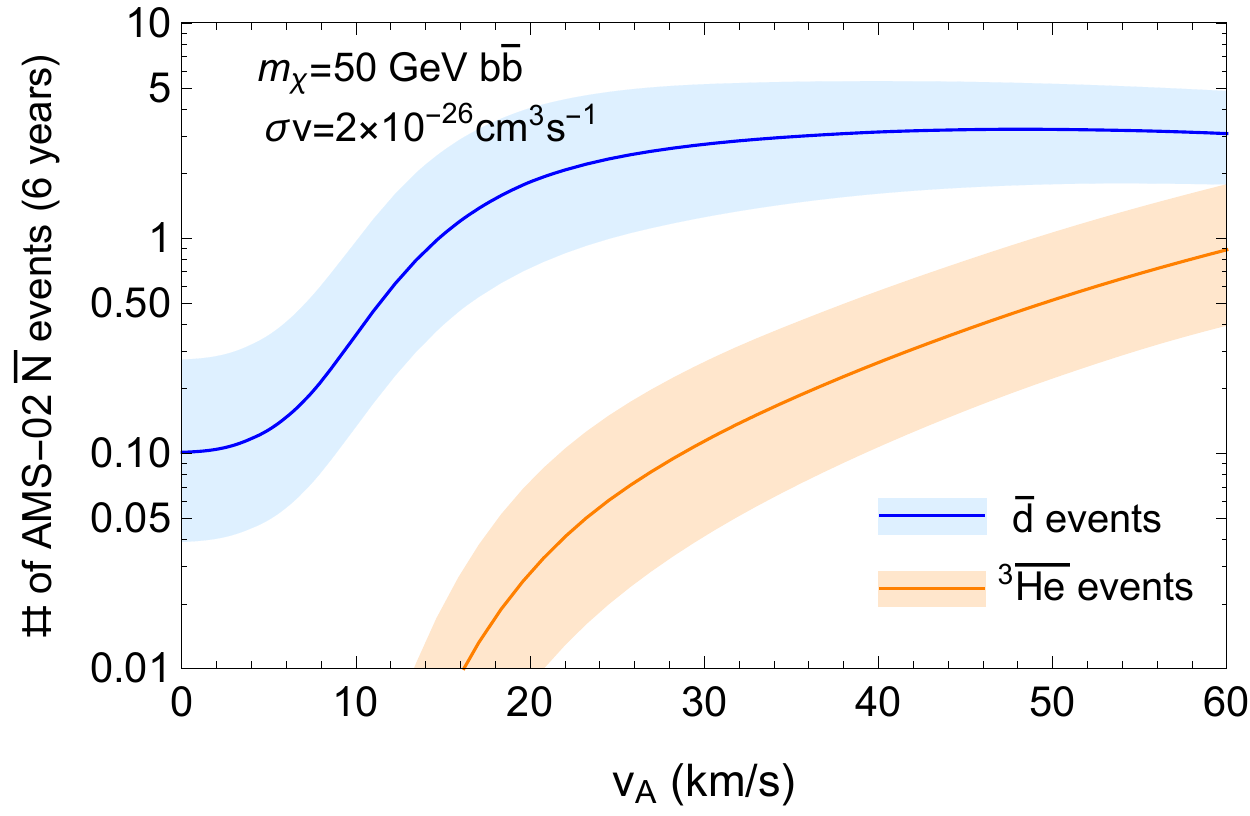}
\includegraphics[width=3.41in,angle=0]{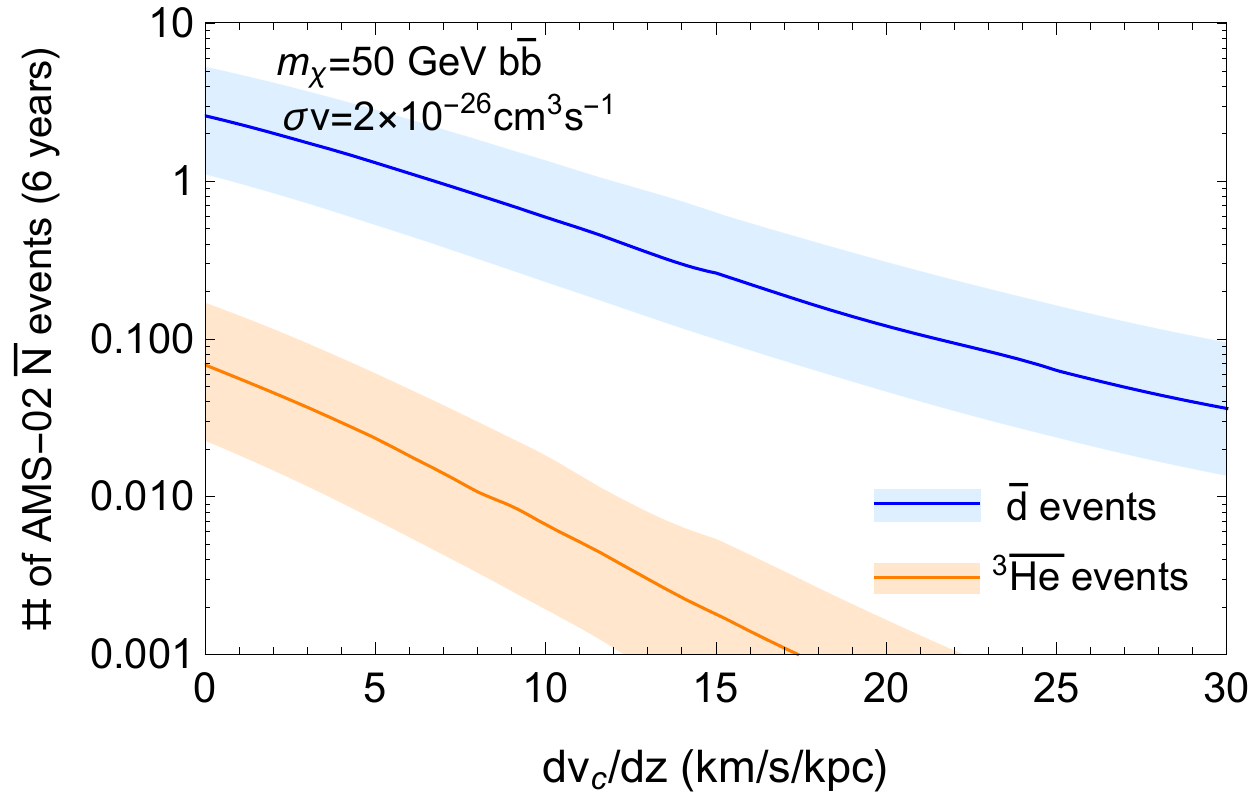}\\
\includegraphics[width=3.41in,angle=0]{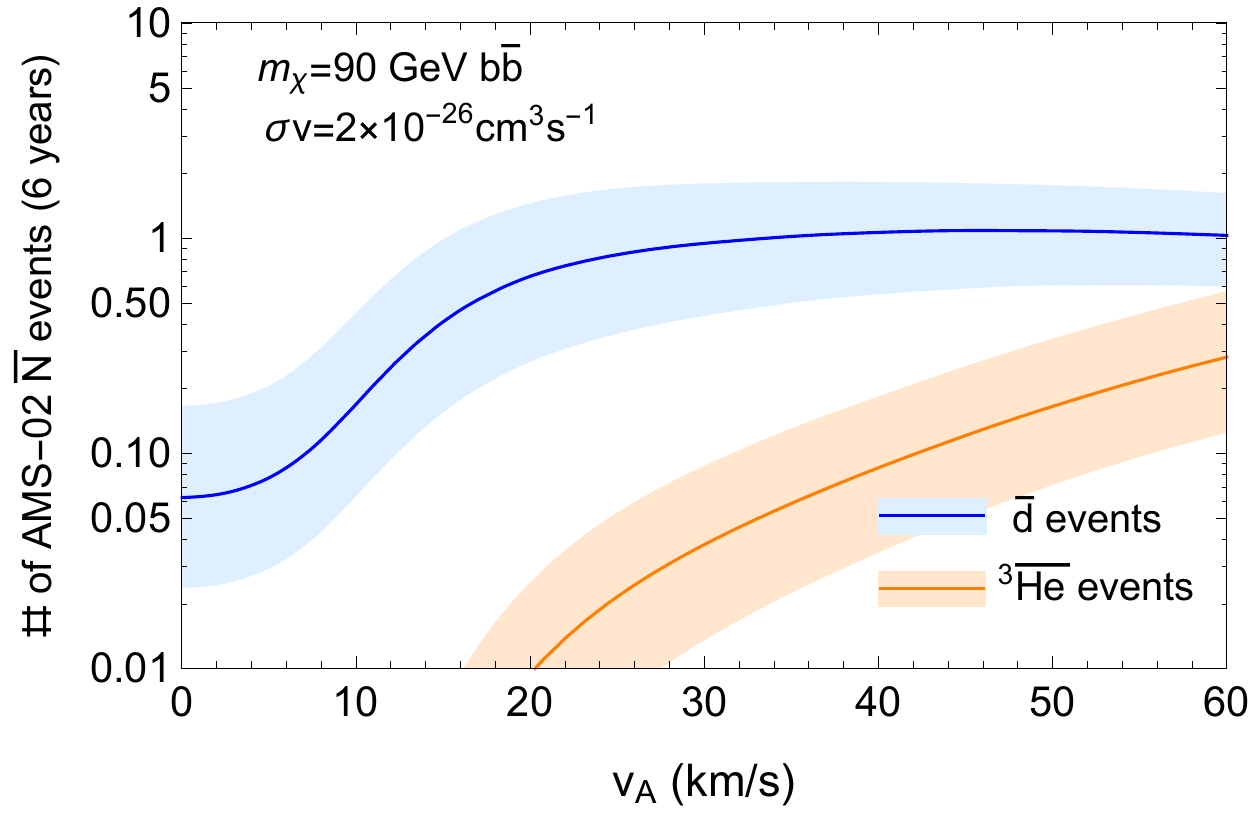}
\includegraphics[width=3.41in,angle=0]{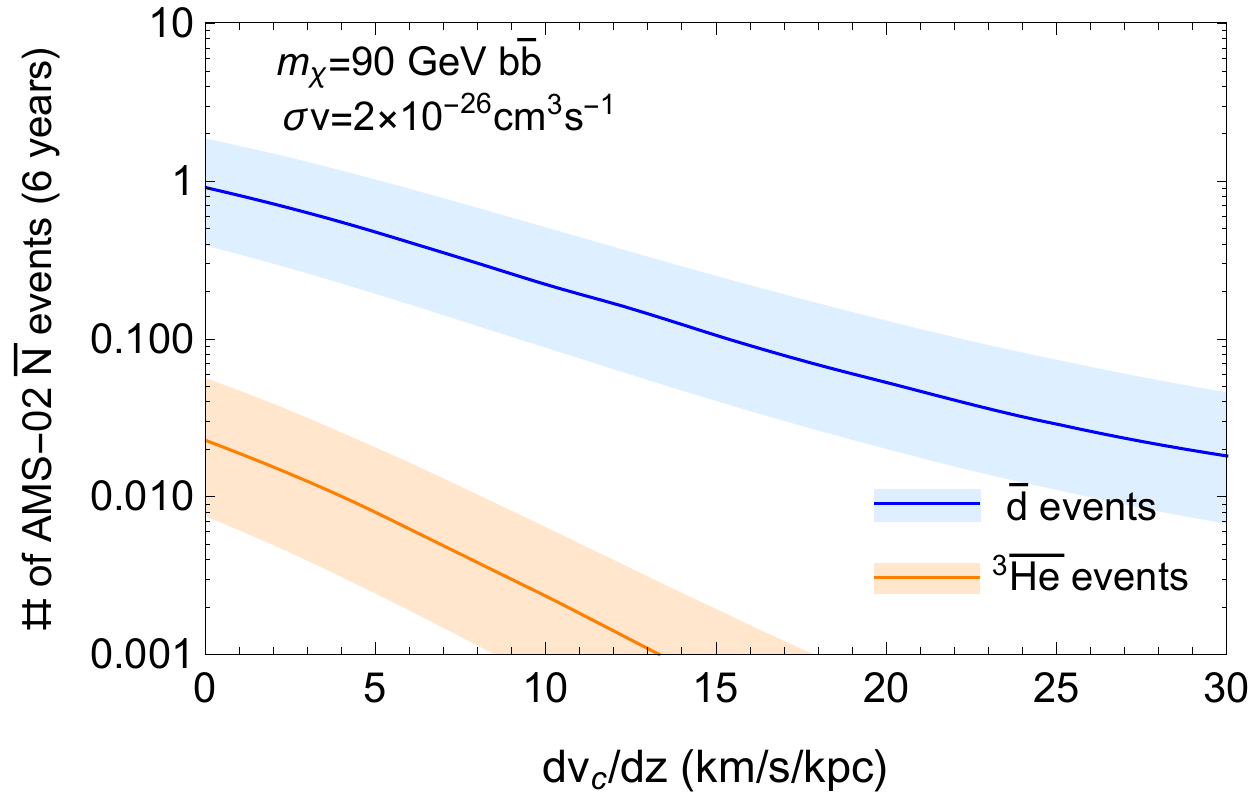}
\caption{As in Fig.~\ref{fig:Convection_and_Reacceleration}, but for dark matter particles with a mass of 50 
GeV (top) or 90 GeV (bottom), and that annihilate to $b\bar{b}$ with a cross section of $\sigma v =  2 \times 10^{-26}$ cm$^3/$s.}
\label{fig:Con_and_Reac_Masses}
\end{figure*}

Increasing the value of the coalescence momentum, $p_{0}$, opens the phase space and leads to larger fluxes of anti-nuclei. This effect is particularly important in the case of heavier anti-nuclei. The impact of the uncertainty in $p_0$ on the local ratio of $\barHet$ and $\bar{\rm d}$ has been explored in Refs.~\cite{Carlson:2014ssa, Cirelli:2014qia, Korsmeier:2017xzj}. Here, we follow Ref.~\cite{Korsmeier:2017xzj}, which includes separate treatments of $p_0$ from dark matter and astrophysical interactions. For anti-nuclei from dark matter annihilation, we adopt $p_{0} = 160 \pm 19$ MeV, based on measurements at e$^+$e$^-$ colliders~\cite{Schael:2006fd}, while for secondary production, we use the range of $p_{0}=$~\mbox{208--262} MeV for $\bar{\rm d}$ and $p_{0}=$~\mbox{218--261} MeV for $\barHet$, based on measurements of proton-proton collisions~\cite{Acharya:2017fvb}. There are no existing measurements for the case of $\barHef$ production, so we adopt the same $p_{0}$ range as we did for $\barHet$.

In addition to the uncertainty associated with the value of $p_{0}$, the inclusive antiproton and antineutron production cross sections in proton-proton collisions are uncertain at a level between $\pm 10\%$ and $\pm 50 \%$ for cosmic-ray daughter particles with rigidities between 0.5 and 500 GV. Combining these factors leads to an overall uncertainty in the anti-deuteron production cross section that is a factor of $\sim$2.5 at 10 GeV and $\sim$8 above 200 GeV. The uncertainties are even larger for $\barHet$ and $\barHef$.\footnote{Annihilations of anti-nuclei provide only a very small correction, at one part in $\sim 10^{3}$. Additionally, the tertiary components of $\bar{\rm d}$, $\barHet$ and $\barHef$ can be absorbed into the uncertainties associated with propagation through the ISM.}

In Fig.~\ref{fig:AntiDeuteronsFluxes}, we show the spectrum of cosmic-ray anti-deuterons predicted from standard astrophysical production, along with the various uncertainties associated with this prediction. The black curve represents our central prediction for the case of ISM Model~I (taking $\chi \chi \rightarrow b \bar{b}$), and adopting the antiproton production cross section of Ref.~\cite{Tan:1983de}, a coalescence momentum of $p_{0} = 262$ MeV, and the best-fit solar modulation model of Ref.~\cite{Cholis:2015gna}. The colored bands represent the uncertainties associated with the effects of solar modulation (green)~\cite{Cholis:2015gna}, the injection and ISM transport model (blue)~\cite{Cholis:2019ejx}, and the antiproton production cross section and coalesence momentum (orange)~\cite{Acharya:2017fvb, galprop}. The larger red band depicts the total uncertainty associated with the combination of these factors.

\section{Dependence on the Dark Matter Mass}
\label{app:A}

In the main body of this paper, we focused on the case of dark matter particles with a mass of 67 GeV. In this section, we show results for two other values of this quantity, 50 and 90 GeV. Changing the dark matter mass can affect the observed number of $\bar{\rm d}$ and 
$\barHet$ events, as shown in Fig.~\ref{fig:Con_and_Reac_Masses}. In the mass range of 20-100 GeV, lighter dark matter particles produce larger numbers of $\bar{\rm d}$ and $\barHet$ events at \textit{AMS-02} (for a given value of $\langle \sigma v \rangle$) due to their higher annihilation rate in the halo of the Milky Way.

\section{Prospects for the GAPS Experiment}
\label{app:A}

\begin{figure}[t!]
\includegraphics[width=3.45in,angle=0]{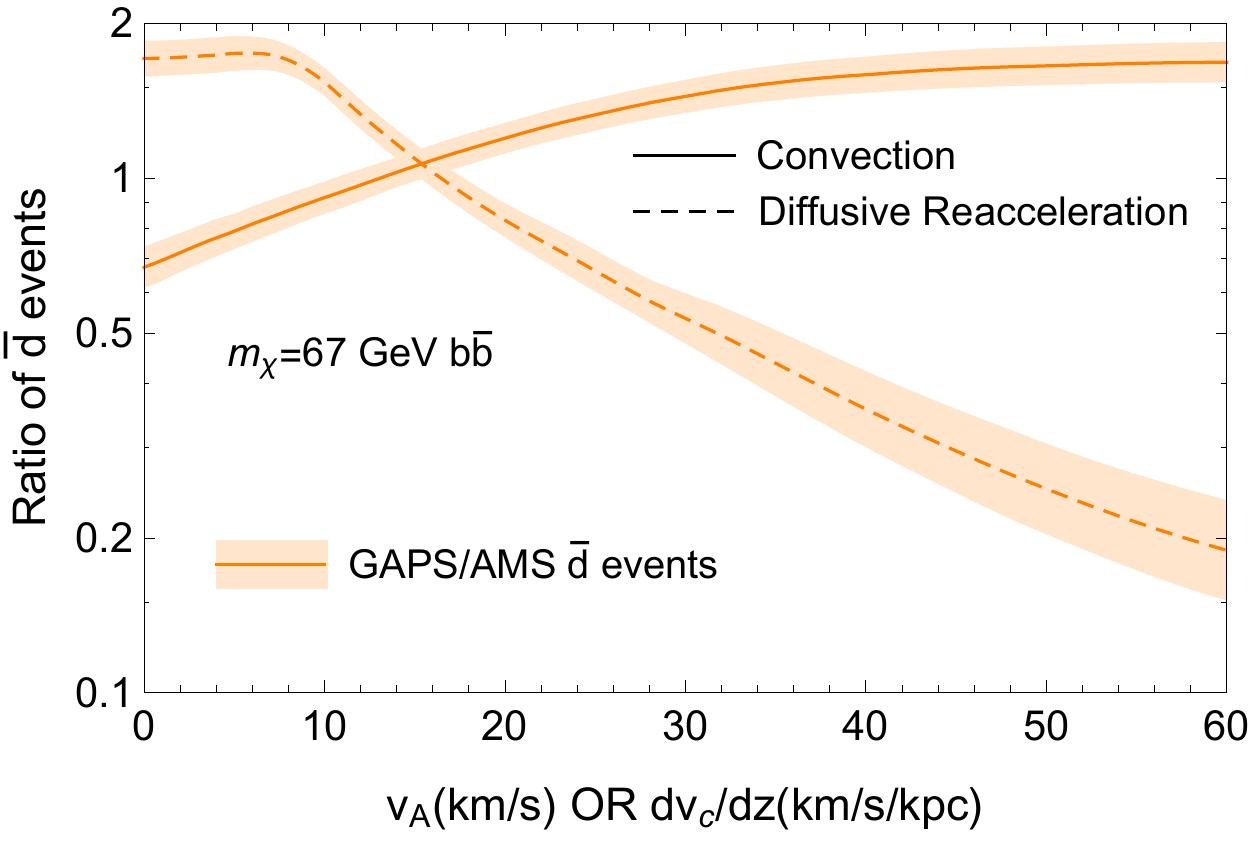}         
\caption{The expected ratio of anti-deuteron events at GAPS to at \textit{AMS-02}, as a function of the Alfv{\'e}n speed and convection velocity. We assume 35 days and 6 years of data from GAPS and \textit{AMS-02}, respectively.}
\label{fig:GAPS_and_AMS}
\end{figure}

Here, we discuss the additional information that could be provided by the General AntiParticle Spectrometer (GAPS) experiment~\cite{Aramaki:2015laa, Lowell:2018xff}. In order to illustrate the complementary of GAPS and \textit{AMS-02}, we plot in Fig.~\ref{fig:GAPS_and_AMS} the predicted ratio of anti-deuteron events at GAPS to that at \textit{AMS-02}, each as a function of the Alfv{\'e}n speed or convection velocity. In calculating these results, we have assumed 35 days and 6 years of data for GAPS and \textit{AMS-02}, respectively. For low values of $v_A$ or high values of $dv_c/dz$, GAPS is expected to observed a larger number of anti-deuteron events than \textit{AMS-02}. 

\end{appendix}

\end{document}